\documentclass[12pt]{article}
\usepackage{color}
\usepackage{amsfonts}
\usepackage{amsthm}
\usepackage{amssymb,bm}
\usepackage{amsmath,enumerate,rotate}
\usepackage{amssymb,latexsym,mathrsfs}
\usepackage{graphicx}

\usepackage{bm}
\usepackage{bbm}
\usepackage{natbib}
\usepackage{graphicx}
\usepackage[font=footnotesize]{caption}
\usepackage[figuresleft]{rotating}
\usepackage{enumerate}
\usepackage{threeparttable}
\usepackage{multirow}
\usepackage{float}  
\usepackage[left=2.5cm,top=2.5cm,right=2.5cm,bottom=2.5cm,bindingoffset=0.5cm]{geometry}
\newtheorem*{algo}{\bm{$BC_{bl}$}}

\newcounter{savecntr}
\newcounter{restorecntr}
\author{Weixuan Zhu \setcounter{savecntr}{\value{footnote}}\thanks{Departamento de Estadistica, Universidad Carlos III de Madrid, Spain} ,
J. Miguel Mar\'{i}n \setcounter{restorecntr}{\value{footnote}}%
  \setcounter{footnote}{\value{savecntr}}\footnotemark
  \setcounter{footnote}{\value{restorecntr}} \, and\,
Fabrizio Leisen \thanks{School of Mathematics, Statistics and Actuarial Science, University of Kent, Canterbury CT2 7NF, U.K.}
}
\begin{document}

\title{A Bootstrap Likelihood approach to Bayesian Computation}

\date{}

\maketitle

\abstract{
There is an increasing amount of literature focused on Bayesian computational methods to address problems with intractable likelihood. One approach is a set of algorithms known as \textit{Approximate Bayesian Computational} (ABC) methods. One of the problems of these algorithms is that the performance depends on the tuning of some parameters, such as the summary statistics, distance and tolerance level.
To bypass this problem, \cite{Robert13} introduced an alternative method based on empirical likelihood, which can be easily implemented when a set of constraints, related to the moments of the distribution, is known.
However, the choice of the constraints is sometimes challenging. To overcome this problem, we propose an alternative method based on a bootstrap likelihood approach. The method is easy to implement and in some cases it is faster than the other approaches. The performance of the algorithm is illustrated with examples in Population Genetics, Time Series and Stochastic Differential Equations. Finally, we test the method on a real dataset.}

\section{Introduction}

Markov chain Monte Carlo (MCMC) algorithms have played an
important role in spreading the adoption of Bayesian methodology to a wide range of applications.
However, modern “big data” applications require complex models and demanding computational techniques; in some of these situations, MCMC methods may be extremely slow or even impossible to implement. These problems naturally arise in many research areas, for example in Population Genetics (\cite{Beaumont02}, \cite{Drovandi10}), Epidemics (\cite{McKinley09}) and Hidden Markov Models (\cite{Dean14}). Recently, a new class of algorithms, called Approximate Bayesian Computational (ABC) methods, have been proposed to evaluate likelihoods which are analytically infeasible or computationally expensive. As remarked by \cite{Robert2012}, the first genuine ABC algorithm was introduced by \cite{Pritchard} in a Population Genetics setting. Precisely, suppose that the data $\bm{y}\in\mathcal{D}\subset\mathbb{R}^n$ is observed. Let $\varepsilon>0$ be a tolerance level, $\eta$ a summary statistic on $\mathcal{D}$ (which often is not sufficient) and $\rho$ a distance on $\eta(\mathcal{D})$. Then, the algorithm works as follows 

\bigskip

\noindent \textbf{for} $i=1$ \textbf{to} $N$ \textbf{do}\\
--{\bf Repeat}\\
----Generate $\theta'$ from the prior distribution $\pi(\cdot)$\\
----Generate $\bm{z}$ from the likelihood $f(\cdot|\theta')$\\
--{\bf until $\rho\lbrace \eta(\bm{z}),\eta(\bm{y})\rbrace\leq \varepsilon$}\\
set $\theta_i=\theta'$\\
\textbf{end for}

\bigskip
 
The basic idea behind the ABC is that, for a small (enough) $\varepsilon$ and a representative summary statistic, we can obtain a reasonable approximation of the posterior distribution. 
Therefore, the choice of a summary statistics, a distance and a tolerance level play a crucial role in implementing an efficient ABC algorithm. In order to relax some of the tuning problems, several recent papers have focused on strategies for setting the parameters of ABC algorithms, see for instance \cite{Fearnhead12}, \cite{Moral12} and \cite{Sisson07}.

Recently there has been a growing interest in methods where approximated likelihoods are used to deal with intractability. For example, 
\cite{Robert13} proposed an alternative approach that uses the well-established empirical likelihood approximation ($\mbox{BC}_{el}$ sampler). The authors apply the method in a Bayesian framework to avoid the choice of the ABC parameters. The basic algorithm works in the following way: Firstly, generate $M$ parameters $\theta_i$ from the prior distribution. Then, set the weight $\omega_i = L_{el}(\theta_i|y)$, where $L_{el}(\theta_i|y)$ is the empirical likelihood of $\theta_i$ given the observed data $y$. The output of $\mbox{BC}_{el}$ is a sample of size $M$ of parameters with associated weights, which operates as an \textit{importance} sampling output. However, the validation of the empirical likelihood depends on the choice of a set of constraints that ensures convergence. 

Our work is in the vein of \cite{Robert13}. In particular, we propose to use the bootstrap likelihood (\cite{Davison92}) approximation instead of empirical likelihood. As for the empirical likelihood, the bootstrap likelihood converges to the true likelihood as the number of observations increases. 

The main motivation of the method is that it does not require setting any objective constraints.
In some problems, for example random field models (see Section 5.4), it is not clear how to set these constraints (see \cite{Owen01}).

The only drawback with this method is that parameter estimators  must be available. Nonetheless, in the population genetics example (Section 5.3) we will show that estimators that arise from approximated likelihoods could lead to good estimation results. In this particular case, composite likelihoods have proved consistent for estimating some parameters such as recombination rates (see \cite{XR2011}).

The outline of the paper is as follows: Section 2 recalls the empirical likelihood approximation and the recent proposal of \cite{Robert13}. The basic bootstrap likelihood method, introduced by \cite{Davison92}, is summarized in Section 3. Section 4 is devoted to the description of our methodology and in Section 5 the methodology is tested on several examples such as Time Series, Population Genetics, Stochastic Differential Equations and Random Fields. 

Throughout the paper, we will denote the algorithm of \cite{Robert13} as $\mbox{BC}_{el}$ and our algorithm as $\mbox{BC}_{bl}$.

\section{Empirical Likelihood and the $\mbox{BC}_{el}$ sampler}

The empirical likelihood has been developed by \cite{Owen01} as a non-parametric version of classical likelihood techniques. The
main ideas of empirical likelihood can be shortly summarized as follows: assume that the dataset $\bm{y}$ is composed of $n$ independent replicates $\bm{y}=\{y_{1},\ldots ,y_{n}\}$ of some random vector $Y$ with density $f$. Rather than defining the likelihood from the density $f$ as usual, the empirical likelihood approach starts by defining the parameters of interest, $\bm{\theta}$, as functionals of $f$, for instance as moments of $f$, and it then profiles a non-parametric likelihood.   
More precisely, suppose that are available a set of constraints of the form
$$\mathbb{E}_f[h(Y,\bm{\theta})]=0,$$
where $h$ is a smooth function of the data and the dimension of $h$  equals the number of constraints that defines $\theta$. The \textit{empirical likelihood} is defined as 
\begin{equation*}
L_{el}(\bm{\theta}|\bm{y})=\underset{p_{i} }{\max }\prod_{i=1}^{n}p_{i}
\end{equation*}
subject to the constraints
\begin{equation*}
\begin{array}{ccc}
\sum_{i=1}^{n}p_{i}=1, &  & \sum_{i=1}^{n}p_{i}h(y_{i};\theta )=0.
\end{array}%
\end{equation*}
For instance, in the one-dimensional case when $\theta=\mathbb{E}_f[Y]$, the empirical likelihood in $\theta$ is the maximum of the product $\prod_{i=1}^{n}p_{i}$ under the constrain $$p_1y_1+\cdots+p_ny_n=\theta,$$ 
since $h(y,\theta)=y-\theta$. 
%
\cite{Robert13} developed an alternative approach to standard ABC methods, based on the empirical likelihood approximation ($\mbox{BC}_{el}$ sampler). The procedure can be summarized as follows
%
%
%
%
%
%
%
%

\begin{minipage}[thp]{330pt} 
\label{alg1}
\par\vspace{5pt}\hrule
\bigskip
\textbf{for} $i=1$ \textbf{to} $M$ \textbf{do}
\begin{enumerate}
\item Generate $\bm{\theta_i}$ from the prior distribution $\pi(\cdot)$
\item Set the weight $w_i=L_{el}(\bm{\theta_i}|\bm{y})$
\end{enumerate}
\textbf{end for}\\
\hrule\vspace{5pt}
\end{minipage}

\noindent where $L_{el}(\bm{\theta_i}|\bm{y})$ is the empirical likelihood of $\bm{\theta_i}$ given the observed data $\bm{y}$.
The output of $\mbox{BC}_{el}$ is a sample of size $M$ of parameters with associated weights, which operates as an importance sampling output.

The main advantages of $\mbox{BC}_{el}$, when compared with standard ABC, are that they neither require simulations from the sampling model, nor any choice of parameters such as summary
statistics, distance measure and tolerance. Bypassing model simulations sometimes leads to significant time savings in complex models, like those found in population genetics. However, $\mbox{BC}_{el}$ still requires delicate calibrations in most cases. In this paper we propose to replace the empirical likelihood in the $\mbox{BC}_{el}$ sampler with a bootstrap likelihood approximation. The main motivation is that the choice of the constraints which ensures the convergence of the empirical likelihood in terms of theorem 3.4 of \cite{Owen01} (as an extension of Wilk's theorem), is sometimes unclear. 

In the next Section, the Bootstrap likelihood approximation and its properties are recalled. 

\section{Bootstrap Likelihood}
\cite{Davison92} use nested bootstrap calculation in conjunction with kernel
smoothing methods to calculate estimates of the density of a given statistic
for a range of parameter values. These density estimates are used to
generate values of an analogue of a likelihood function by curve-fitting
methods (see also Efron and Tibshirani (1994), and Davison and Hinkley
(1997)).

Assume that $\hat{\theta}$ is an estimator of a parameter of interest $%
\theta $ and we seek an approximate likelihood function for $\theta .$ The
goal is to estimate the sampling density $p\left( \hat{\theta}|\theta
\right) ,$ namely, the the sampling distribution of $\hat{\theta}$ when the
true parameter is $\theta .$ The basic method can be summarized as follows:

\begin{itemize}
\item Suppose $\theta $ is the parameter of interest and $\hat{\theta}$ is
the parameter estimated by its sample analogue. Generate $K$ bootstrap
samples of size $n$ (same size as the original data) to obtain a series of
populations $P_{1}^{\ast },\ldots,P_{K}^{\ast }$ giving bootstrap replications $%
\hat{\theta}_{1}^{\ast },\ldots,\hat{\theta}_{K}^{\ast }$ (first-level
bootstrap). Any estimation method can be used apart from likelihood
estimators.

\item For each of the $i$-th bootstrap samples $P_{i}^{\ast }$ we generate $%
L $ samples of size $n$ (where $L$ is preferably 1000, as suggested in \cite{Davison92}). For each sample calculate the analogue of $\theta $, denoted
by $\hat{\theta}_{ij}^{\ast \ast }$ (second-level bootstrap) giving the
second stage bootstrap replicates. We form kernel density estimates at each point $\hat{\theta}_{i}^{\ast }$%
\[
\hat{p}\left( t|\hat{\theta}_{i}^{\ast }\right) =\frac{1}{L\cdot s}%
\sum_{j=1}^{L}\ker \left( \frac{t-\hat{\theta}_{ij}^{\ast \ast }}{s}\right)
\]%
for $i=1,\ldots ,K.$ In this case $\ker (\cdot )$ is any kernel function. We then evaluate $\widehat{p}\left( t|\hat{\theta}_{i}^{\ast }\right) $ for $%
t=\hat{\theta}.$ Since the values $\hat{\theta}_{ij}^{\ast \ast }$ were
generated from a distribution governed by parameter value $\hat{\theta}%
_{i}^{\ast }$ then $\widehat{p}\left( \hat{\theta}|\hat{\theta}_{i}^{\ast
}\right) $ provides an estimate of the likelihood of $\hat{\theta}$ for parameter
value $\theta =\hat{\theta}_{i}^{\ast }.$ Then the $K$ values $l(\hat{\theta}_{i}^{\ast})=\log [\widehat{p}\left( \hat{\theta}|\hat{\theta}_{i}^{\ast
}\right) ]$ are obtained.

\item Apply a smooth curve-fitting algorithm, like a scatterplot smoother
to the pairs $\left( \hat{\theta}_{i}^{\ast },l(\hat{\theta}_{i}^{\ast})\right) $
for $i=1,\ldots ,K,$ to obtain the whole log bootstrap likelihood curve.
\end{itemize}

Although the previous scheme is adapted to the case of \emph{i.i.d.} samples, in the case of dependent data, such as regression-type problems, the outlined method also applies. There are different ways to specify the bootstrap data generating processes for dependent data, while some involve strong assumptions, others very weak ones (see \cite{Davison06}).

We may consider among them, pairs bootstrap or pairwise bootstrap. The idea is simply to resample the data, keeping the dependent and independent variables together in pairs. It can be applied to a large range of models.

Another main approach  is called the residual bootstrap (residual resampling), which requires that the errors to be \emph{i.i.d.} and independent of regressors, but with minimal distributional assumptions. 
Other possible alternatives are the parametric bootstrap, which assumes that the error terms follow a known distribution, and the wild bootstrap, specifically designed to handle heteroskedasticity in regression models.

From a practical point of view, compared to residual bootstrap and wild bootstrap, pairs bootstrap yields in general less accurate results (see \cite{Mackinnon06}).

The relation between empirical likelihood and bootstrap likelihood is also
explored from a theoretical point of view in \cite{Davison92} and \cite{Owen01}. They point out that the bootstrap likelihood matches the empirical
likelihood to first order. Specifically, in the case of an estimator determined by a monotonic
estimating function, standardized so that the leading term is of order one,
it is shown by applying empirical cumulants (see \cite{Davison92}) that
empirical and bootstrap likelihoods agree to order $n^{-\frac{1}{2}}$ but
not to order $n^{-1}$ in the number of observations, $n$. In this way,
results derived for empirical likelihood, such as the analogue of Wilks'
theorem, apply also to the bootstrap likelihood (see \cite{Davison92}).

\begin{figure}[ht]
\centering
\includegraphics[width=8cm,height=4cm]{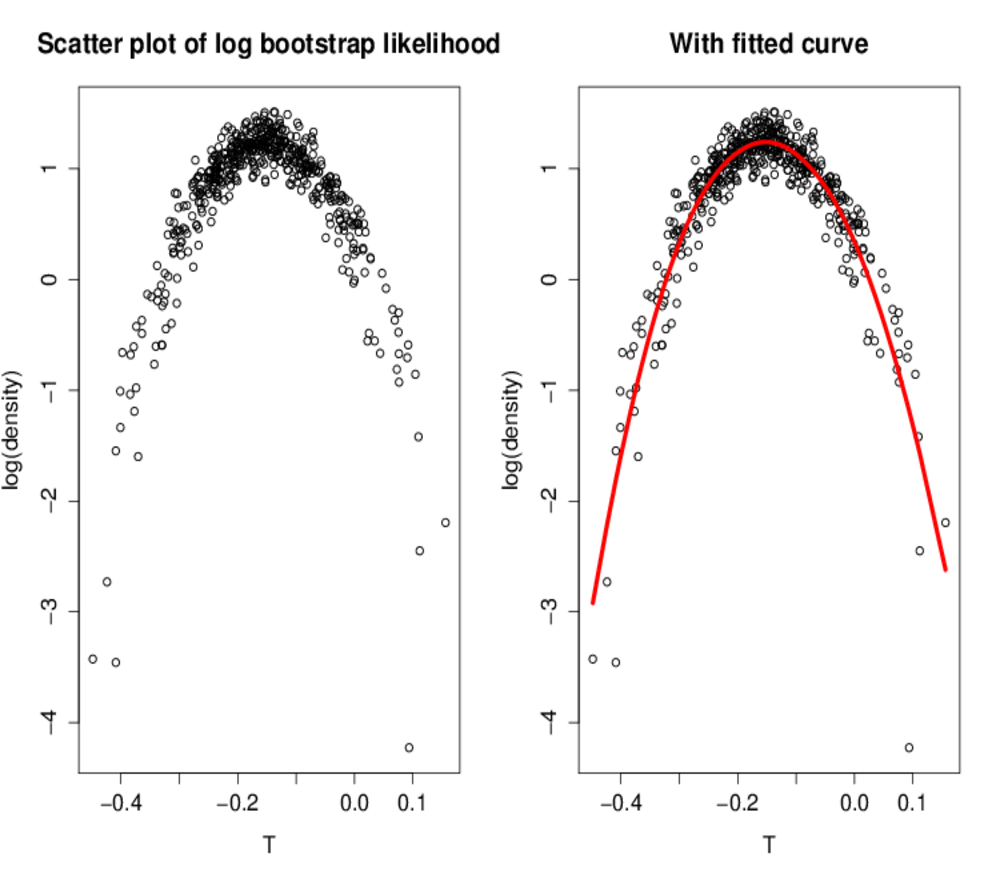}
\caption{\footnotesize{The left figure is plotted after the first two steps in the above
summarized method. Basically, the first-level bootstrap is for generating
the x-axis values and the second-level bootstrap is for the estimation of
the density of $T$ at these corresponding x-axis values. The right figure
displays the estimated bootstrap likelihood curve.}}\label{1}
\end{figure}

Figure \ref{1} is an illustration of the bootstrap likelihood construction. We have used here, and throughout the paper as the smooth curve-fitting algorithm, a local polynomial procedure by means of the \texttt{R} command \texttt{loess}.
In the next section, we will use the bootstrap likelihood to develop an
algorithm to address Bayesian inference in the spirit of \cite{Robert13}.

\section{Bayesian computation via bootstrap likelihood}

Let $BL(\bm{\theta_i}|\bm{y})$
denote the estimation of the bootstrap likelihood in the point $\theta_i$ given the observed data $\bm{y}$. Our sampler works as follows

\begin{center}
\begin{minipage}[thp]{330pt} \par\hrule\vspace{5pt}
\begin{algo}{Bayesian Computation with bootstrap likelihood}
\label{alg1}
\par\vspace{5pt}\hrule
\bigskip
Estimate the bootstrap likelihood curves of parameters with the samples described in the previous section.\\
\textbf{for} $i=1$ to $M$ do
\begin{enumerate}
\item Generate $\theta_i$ from the prior distribution $\pi(\cdot)$
\item Set the weight $w_i = BL(\bm{\theta_i}|\bm{y})$
\end{enumerate}
\textbf{end for}\\
\end{algo}
\hrule\vspace{5pt}
\end{minipage}
\end{center}

The output is a sample of size $M$ of parameters with associated weights, which operate as an importance sampling output. This means that a posterior sample of simulated parameters of size $N$ is sampled with replacement from the $M$ parameters with corresponding weights $w_{i}$'s. The bootstrap likelihood approach allows us to define an algorithm with the same structure of
the one defined in \cite{Robert13}. In contrast with the empirical likelihood method, the bootstrap likelihood doesn't require any set of subjective constraints by virtue of the bootstrap likelihood methodology. 

Another benefit of using the bootstrap likelihood instead of the empirical likelihood is that the construction of bootstrap likelihood does not depend on the priors. Once the bootstrap likelihood curve is fitted (last step of constructing the bootstrap likelihood), the weight $w_i$ in $\mbox{BC}_{bl}$ sampler is obtained directly by taking values on the fitted curve. In contrast, the $\mbox{BC}_{el}$ sampler requires solving an optimization problem at each iteration.
This
leads to significant gain in the computing time when different priors
are compared. At the same time, we have to point out that the same approach can be also realized with the empirical likelihood setting when a very large collection of likelihood values has been gathered.  

\begin{figure}[ht]
\centering
\includegraphics[width=10cm,height=6cm]{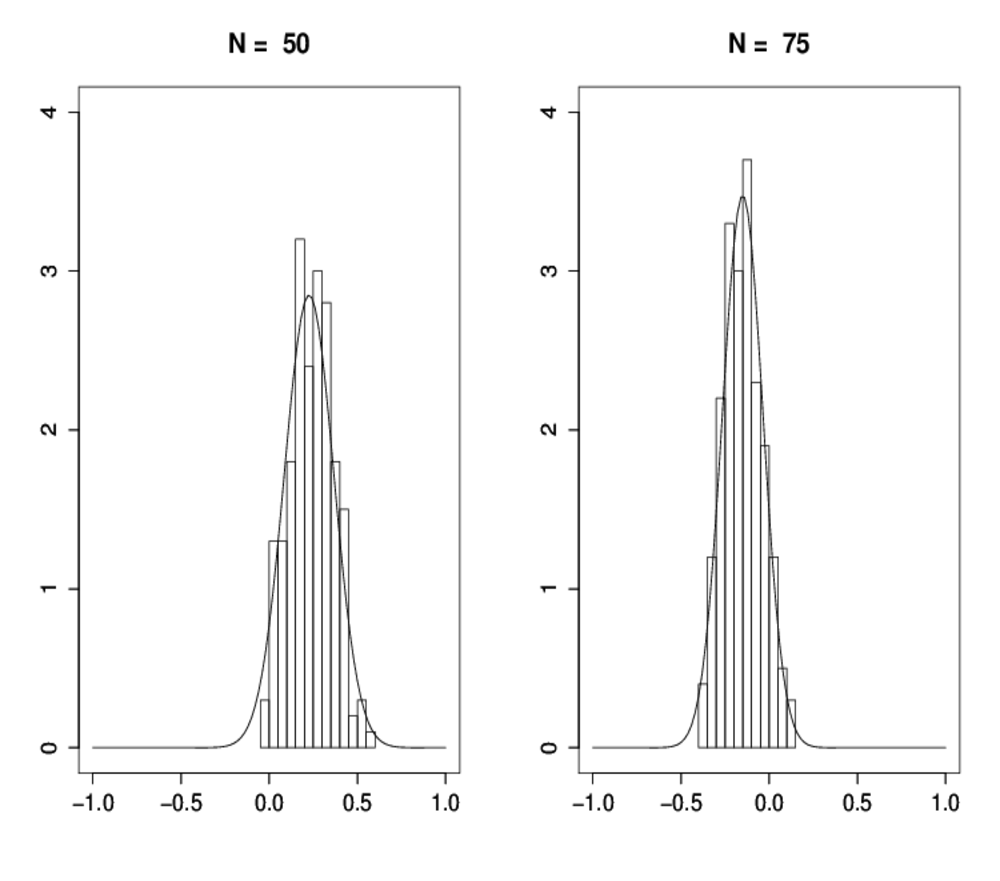}
\caption{\footnotesize{Comparison of the true posterior on the normal mean (solid lines)
with the empirical distribution of weighted simulations resulting from $
\mbox{BC}_{bl}$ algorithm. The normal sample sizes are 50 and 75
respectively, the number of simulated $\protect\theta$'s is 200. }}
\label{Figure_2}
\end{figure}

As a toy illustration of the method, we apply the $\mbox{BC}_{bl}$ algorithm
to a normal distribution with known variance (equal to 1). Clearly, the
parameter of interest is $\mu$, i.e. the unknown location parameter.  We can see in Figure \ref{Figure_2} the fitting of the posterior distribution. In this experiment, the computing time of $\mbox{BC}_{bl}$ algorithm is much less than $\mbox{BC}_{el}$ method. The main reason is that the estimation of $\mu $ (sample mean in this case) is explicit and straightforward, without need for numerical estimation algorithms.

In the next Section, the performance of the bootstrap likelihood approach is explored in several examples. In particular, we will see how to manage the parameter estimation in the nested bootstrap. As we will see, this step of the methodology can vary with the problem at the hand.

\section{Numerical Illustration}\label{Example}

\subsection{Dynamic Models} \label{ex1}

As mentioned in Section 3, one way to deal with the dependence in dynamic models, is through the application of the bootstrap procedure to the unobserved i.i.d. residuals. For example, we test the GARCH(1,1) model: 
$$y_t = \sigma_t\epsilon_t, \ \ \ \epsilon_t\sim N(0,1),\ \ \ \sigma_t^2=\alpha_0+\alpha_1y_{t-1}^2+\beta_1\sigma_{t-1}^2$$
under the constraints $\alpha_0,\alpha_1,\beta_1>0$ and $\alpha_1+\beta_1<1$ (see \cite{Bollerslev86}). An exponential $Exp(1)$ and a Dirichlet $Dirich(1,1,1)$ prior distributions are assumed, respectively, on $\alpha_0$ and $(\alpha_1,\beta_1,1-\alpha_1-\beta_1)$. In order to compare with $\mbox{BC}_{el}$, we set the constraints for the empirical likelihood as in \cite{Robert13}. The respective number of first and second level of bootstrap replicates are $K=100$ and $L=1000$. For each bootstrap replicate, the \texttt{R} function \texttt{garch} from \texttt{tseries} package is used for the estimation of the parameters. This package uses a Quasi-Newton optimizer to find the maximum likelihood estimates of the conditionally normal model. The \texttt{garch} function provides a fast estimation of the parameters but it does not always converge consistently. Another alternative may be using the \texttt{garchFit} function from \texttt{fGarch} package that is slower but converges better.

\begin{figure}[H] 
\centering
\includegraphics[width=8cm,height=6cm]{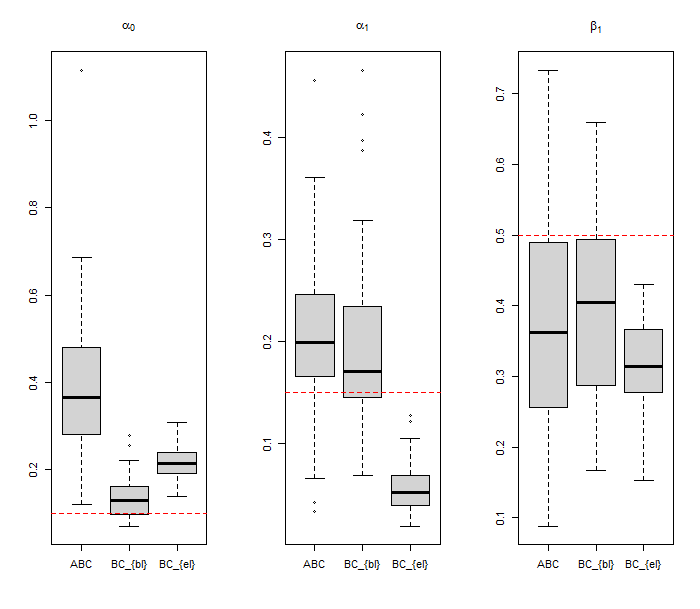}
\caption{\footnotesize{Comparison of evaluations of posterior expectations. (with true values in dashed lines) of the parameters $(\alpha_0, \alpha_1, \beta_1)$ of the $GARCH(1,1)$ model with 300 observations.\label{3}}}
\end{figure}

Despite the lack of stability of the \texttt{garch} function, in Figure \ref{3} we can see that the $\mbox{BC}_{bl}$ algorithm is performing better than the $\mbox{BC}_{el}$ algorithm in terms of the ability to find the correct range of $\alpha_0$, $\alpha_1$ and $\beta_1$. Furthermore, Table \ref{table:garch} illustrates that all parameters are accurately estimated with $\mbox{BC}_{bl}$ with small mean square errors (MSE), while the estimations with $\mbox{BC}_{el}$ are poorer in this case.
One potential reason for the poor performance of $\mbox{BC}_{el}$ is the choice of the score constraints for the empirical likelihood adopted by \cite{Robert13}, which might not guarantee its convergence.

\begin{table}[h] \footnotesize
\centering

\begin{tabular}{|c|c|c|c|}
\hline
    True values          & $\mbox{BC}_{bl}$ & $\mbox{BC}_{el}$ & ABC\\ \hline
 $\alpha_0=0.1$ & 0.12886(0.00237) & 0.19782(0.01039) & 0.39454(0.11774)\\ \hline
 $\alpha_1=0.15$ & 0.15307(0.00296) & 0.06277(0.01097) & 0.20551(0.00911)\\
 \hline
 $\beta_1=0.5$ & 0.42874(0.02317) & 0.31218(0.03731) & 0.37449(0.03954)\\
 \hline
\end{tabular}
\caption{\footnotesize{Summaries of the estimates from two approaches. The results are based on 50 simulated datasets, and displayed with true values in the first column, posterior means from $\mbox{BC}_{bl}$ in the second and posterior means from $\mbox{BC}_{el}$ in the last (with MSE reported inside brackets).}}
\label{table:garch}
\end{table}


Finally, from the computational point of view, we note in our experiments that our approach is faster than the empirical likelihood one (1.3 mins vs 2.5 mins).  This is not surprising mainly because the bootstrap likelihood procedure depends heavily on the parameter estimation methodology. In this example, the R function \texttt{garch} provides a quick estimation of the model parameters and consequently a shorter computational time.

\subsection{Stochastic differential equations}\label{SDESEC}

Stochastic differential equations can be used to model random evolution processes along continuous time, e.g. they are commonly used in many applied areas such as financial models, population dynamics or pharmacokinetics studies. Statistical inference for stochastic differential equations has been undertaken usually from a frequentist point of view, although new Bayesian methodologies have been recently proposed (see \cite{picchi14}).

With respect to an empirical likelihood approach to SDE models, there are quite a few references, and the methodology is based on a Kernel estimator, in order to calculate the maximum of the corresponding functional expressions (see \cite{Zhengyan10}). More precisely, the constrains of the empirical likelihood are defined on a set of functionals of Kernel densities. A BC$_{el}$ based on a Kernel approach seems to tangle the procedure, so we think that it is more useful to make the comparison with a standard ABC method.

In this section we focus on an example taken from \cite{Brouste14} and we compare the $BC_{bl}$ procedure with a standard ABC method.

We consider the model 
\begin{equation*}
dX_{t}=(2-\theta_{2}X_{t})dt+\left(1+X_{t}^{2}\right)^{\theta_{1}}dW_{t},
\end{equation*}
where $X_{0}=1$, and we simulate a set of 750 data points assuming $\theta _{1}=0.2$ and $\theta_{2}=0.3$.

We apply first a pure rejection sampling scheme for ABC, with uniform $U(0,1)$ prior distributions for $\theta_{1}$ and $\theta_{2}$. We run 5000 simulations with a tolerance equal to $0.1$; as summary statistics we have addressed the mean, variance and mean absolute deviation (mad). We have used the library \texttt{EasyABC} for computing tasks.

With regard to $BC_{bl},$  we use a parametric bootstrap version where the respective number of the first and second levels of bootstrap replicates are $K=100$ and $L=200$. For each bootstrap replicate we estimate the parameters by means of a quasi maximum likelihood procedure; in this case, we use the function \texttt{qmle} from the \texttt{R} package  \texttt{yuima}.

\begin{table}[h]
\centering
\small
\begin{tabular}{|c|c|c|}
\hline
True values & ABC & $BC_{bl}$ \\ \hline
$\theta_{1}=0.2$ & 0.28644 (0.01300) & 0.20144 (0.00008) \\ \hline
$\theta_{2}=0.3$ & 0.41261 (0.02420) & 0.34773 (0.02360) \\ \hline
\end{tabular}
\caption{\footnotesize{Summaries of the estimates from two approaches. The results are based on 50 simulated datasets, and displayed with true values in the first column, posterior means from ABC in the second and posterior means from $\mbox{BC}_{bl}$ in the last (with MSE reported inside brackets).}}
\label{table:SDE}
\end{table}

In Table \ref{table:SDE} results for both procedures are obtained: estimates by ABC and $BC_{bl}$ are shown with the corresponding \emph{MSE} based in 50 replicates of the model. Here, the estimation of parameters with the ABC approach seems to behave less accurately than $BC_{bl}$, although we have used quite restricted prior distributions to perform the ABC simulations. By using less informative prior distributions, results are still less favourable in the case of the ABC method. Computing times are similar in both cases (7 mins with $BC_{bl}$ and 6 mins with ABC), although it increases dramatically in the case of ABC methods when more iterations are required for a better approximation of the estimates.
\subsection{Population Genetics}\label{PopGen}

ABC methods are very popular in population genetics, see e.g. \cite{Cornuet14}. \cite{Robert13} compare the performance of the $\mbox{BC}_{el}$ sampler with a traditional ABC in the context of evolutionary history of species. They showed that the results are in favor of $\mbox{BC}_{el}$ both in efficiency and effectiveness.

In this section we focus on the study of the distribution of microsatellites, which are repeating sequences of short base pairs of DNA.
They are used as molecular markers for kinship and fingerprinting. For a given genetic locus we can consider different types of alleles (genes), namely, alternative forms of the same genetic locus.

The main caution when applying bootstrap likelihood in such setting is the choice of parameter estimates inside each bootstrap level. The true likelihood is intractable in most population genetic settings due to the complexity of the models. However, composite likelihoods have been proved consistent for estimating some parameters such as recombination rates (see \cite{XR2011}). We will adopt the maximum composite likelihood estimators as parameter estimates in bootstrap likelihood.

Specifically, the intra-locus likelihood is approximated by a product over all pairs of genes in the sample at a given locus. Let $y_{i}^{k}$ denote  the \emph{i}-th gene at the \emph{k}-th locus and $\phi =(\tau ,\theta )$ the vector of parameters; then the pairwise likelihood of the data at the \emph{k}-th locus, namely $y^{k}$, is defined by
\[
l_{2}\left( y^{k}|\phi \right) =\prod\limits_{i<j}l_{2}\left(
y_{i}^{k},y_{j}^{k}|\phi \right)
\]%
where 
\begin{equation*}
l_{2}\left( y_{i}^{k},y_{j}^{k}|\phi \right) =\left\{ 
\begin{array}{lcl}
\frac{\rho (\theta )^{|y_{j}^{k}-y_{i}^{k}|}}{\sqrt{1+2\theta }} &  & \text{%
if same deme} \\ 
\frac{e^{-\tau \theta }}{\sqrt{1+2\theta }}\sum_{m=-\infty }^{\infty }\rho
(\theta )^{|m|}I_{|y_{i}^{k}-y_{j}^{k}|-m}(\tau \theta ) &  & \text{if
different deme}%
\end{array}%
\right. 
\end{equation*}%
and 
\begin{equation*}
\rho (\theta )=\frac{\theta }{1+\theta +\sqrt{1+2\theta }}.
\end{equation*}

In the above equation, $I_{\delta}(z)$ denotes the $\delta$th-order modified Bessel function of the first kind evaluated at $z$.
Note that the expression of the different \textit{deme} case involves an infinite sum, but in practice only the first few terms are required for an accurate approximation, because the value of $m$ corresponds to the number of pairs of mutations in opposite directions, which is usually very small (see \cite{Wilson98}).

\begin{figure}[tbp]
\centering
\includegraphics[width=4cm,height=3cm]{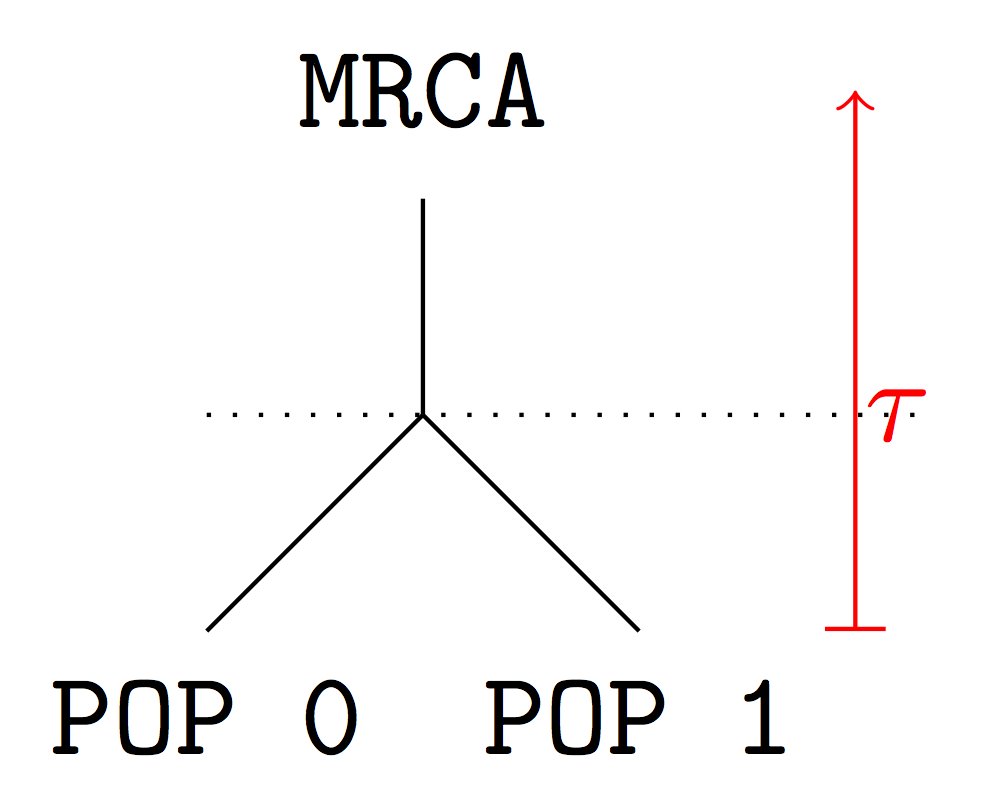}
\caption{\footnotesize{Evolutionary scenario of genetic experiment.}}
\label{4}
\end{figure}

We compare our proposal with $\mbox{BC}_{el}$ in the first evolutionary scenario studied in \cite{Robert13}, see Figure \ref{4}.

Briefly, we simulate data from the genealogy at a given locus until the most recent common ancestor according to coalescence theory. Then a single mutation event is put at random on one branch of the genealogy. In this scenario, there are two parameters of interest $\tau $ and $\theta $. Specifically, $\tau $ is the time at which the two populations diverged in the past and $\theta /2$ is the mutation rate of the mutations at a given locus. The simulated datasets are made of ten diploid individuals per population genotyped at fifty independent loci. We use the \texttt{DIYABC} software (see \cite{Cornuet14}) for simulations of the population. The purpose of this experiment is to analyze it as a test case.

\begin{table}[h]
\centering
\begin{tabular}{|c|c|c|}
\hline
              & $\mbox{BC}_{bl}$ & $\mbox{BC}_{el}$ \\ \hline
 $\theta=10$ & 9.74168(3.76261) & 9.38650(3.35539) \\ \hline
 $\tau=0.5$ & 0.42101(0.02918) & 0.54501(0.13742)\\
 \hline
 
\end{tabular}
\caption{\footnotesize{Summaries of the estimates from two approaches. The results are based on 20 simulated datasets, and displayed with true values in the first column, posterior means from $\mbox{BC}_{bl}$ in the second and posterior means from $\mbox{BC}_{el}$ in the last (with MSE reported inside brackets).}}
\label{table:genetics}
\end{table}
All details about implementations of the $\mbox{BC}_{el}$ procedure can be fully found in \cite{Robert13}. By comparing the posterior means and MSE in Table \ref{table:genetics}, one can find a similar accuracy and precision of the estimates from both $\mbox{BC}_{bl}$ and $\mbox{BC}_{el}$ samplers. We then compare the marginal posterior distributions of the parameters $\theta $ and $\tau $  obtained with both samplers.

\begin{figure}[ht]
\centering
\includegraphics[scale=0.65]{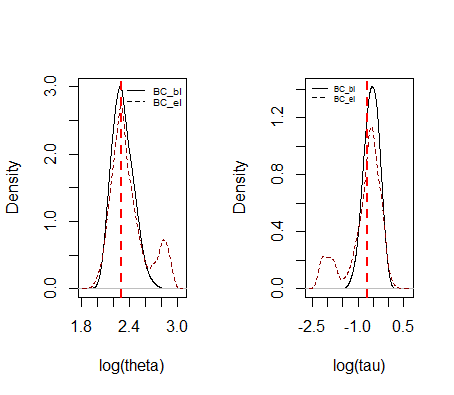}
\caption{\footnotesize{Comparison of the marginal distributions obtained by the $\mbox{BC}_{el}$ and the $\mbox{BC}_{bl}$ sampler. The solid line is the estimated density of samples from $\mbox{BC}_{bl}$ and the dashed line is of samples from $\mbox{BC}_{el}$.}}
\label{5}
\end{figure}

\begin{figure}[ht!]
\centering
\includegraphics[scale=0.65]{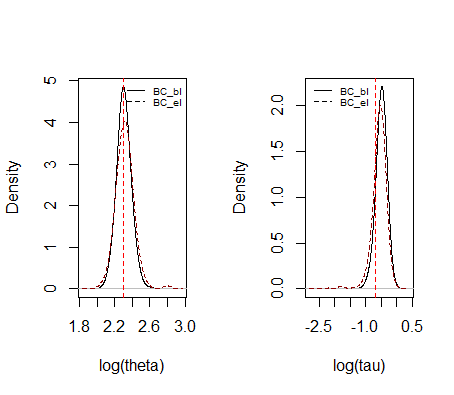}
\caption{\footnotesize{Comparison of the $BC_{el}-\mbox{AMIS}$ and the $BC_{bl}-\mbox{AMIS}$ sampler. The solid line is the estimated density of samples from $\mbox{BC}_{bl}-\mbox{AMIS}$ and the dashed line is of samples from $\mbox{BC}_{el}-\mbox{AMIS}$.}}
\label{6}
\end{figure}

Figure \ref{5} shows samples from the marginal posterior distributions of $\log (\tau)$
and $\log (\theta)$, based on the simulated data. 
Figure \ref{5} suggests that $\mbox{BC}_{el}$ has difficulties eliminating the tails of both posterior distributions and $\mbox{BC}_{bl}$ is more accurate in terms of the shape.
\cite{Robert13} further suggest the incorporation of empirical likelihood in the adaptive multiple importance sampling (AMIS) to speed up the computation. The bootstrap likelihood could also be incorporated in the same way. However, Figure \ref{6} shows that AMIS improves substantially the results computed with the basic $\mbox{BC}_{el}$ sampler, but not so much with respect to the $\mbox{BC}_{bl}$ sampler. For instance, in the case of parameter $\tau$, using AMIS does not improve the performance of $\mbox{BC}_{bl}$ with respect to the true value of the parameter. 
It appears that the basic $\mbox{BC}_{bl}$ sampler is enough capable of building a reasonable posterior, which suggests that it is unnecessary to introduce the AMIS in the bootstrap likelihood setting.

About the computing time, in general, the speed of $\mbox{BC}_{bl}$  depends on many factors, mainly including the numbers of first and second level bootstrap replicates, and the difficulties in estimating the parameter inside each bootstrap level. In this example, the \textbf{R} function \textbf{optim} is employed to estimate the maximum composite likelihood estimator. The speed of $\mbox{BC}_{el}$ depends on the difficulty to optimize under the constraints and  the size of the Monte Carlo sample. For this reason we resort to the \textbf{R} library \textbf{emplik} for the calculation of the empirical likelihood. In this experiment, we also noticed that the computing time of $\mbox{BC}_{bl}$ is more or less twice the time needed for $\mbox{BC}_{el}$ (10 hours vs 4.5 hours) under our parameter setting (50 bootstrap replicates in the first level and 100 replicates in the second for bootstrap likelihood, 30000 Monte Carlo samples in $\mbox{BC}_{el}$).

%

\subsection{Ising and Potts Model}

Ising and Potts models are discrete Gibbs random field models with a statistical physics origin, which are now widely used for applications in spatial modelling, image processing, computational biology, and computational neuroscience.
Consider the simple case of a random field where the pixels of the image $\bm{x}$ can only take two colours (white and black, say). Let $\{\bm{x}=x_{ij}:(i,j) \in D \}$ denote the observed binary data, where $x_{ij}$ is a pixel and $D$ is an $M\times N$ lattice indexing the pixels. The conditional distribution of a pixel is then Bernoulli, with the parameter being a function of the number of neighbouring pixels that have the same value. It is defined as
$$f(x_{ij}=k|x_{n(i,j)})\propto \mbox{exp}(\beta n_{i,j}^k),\ \ \ \ \beta>0,\ \  k=0,1$$
where
$$\ n_{i,j}^k=\sum_{l \in n(i,j)}\mathbbm{l}_{x_l=k}$$
is the number of neighbours of $x_{ij}$ with colour $k$ and $n(i,j)=\{(i+1,j),(i-1,j),(i,j+1),(i,j-1)\}$ is the defined neighbourhood structure. In statistical mechanics, $\beta$ is a strictly positive parameter which can be interpreted as the inverse of the temperature.  The Ising model is defined through these full conditionals
$$f(x_{ij}=1|x_{n(i,j)})=\dfrac{\mbox{exp}(\beta n_{i,j}^1)}{\mbox{exp}(\beta n_{i,j}^0)+\mbox{exp}(\beta n_{i,j}^1)}$$
and the joint distribution therefore satisfies
\begin{equation}
f(\bm{x})\propto \mbox{exp}\left(\beta\sum_{(i,j)\sim (i',j')}\mathbbm{l}_{\{x_{ij}=x_{i'j'}\}}\right) \notag
\end{equation}
where the summation is taken over all the neighbour pairs, namely, a neighbourhood relation on pixels is denoted as $\sim$, where $i\sim j$ denotes that $i$ and $j$ are \emph{neighbours}.
This joint distribution can be obtained from the conditional distributions, by applying the Hammersley–-Clifford representation (see \cite{Grimmett10}).
The Potts model is the natural extension of the Ising Model where more than two colours are considered, see \cite{Robert2014}.

The normalizing constant $Z(\beta)$ of the above distribution depends on $\beta$ and it is numerically tractable only for very small lattices $D$,
which becomes a major obstacle when making inference on $\beta$. The maximum pseudo-likelihood estimator (MPLE) \cite{Besag1977} provides a way to handle the problem. MPLE takes the value that maximizes the pseudo-likelihood function
$$L(\beta|\bm{x})=\prod_{i=1}^M\prod_{j=1}^Nf(x_{ij}=1|x_{n(i,j)},\beta)$$
We will adopt MPLE as the estimation tool to construct the bootstrap likelihood for $\beta$ later. \cite{Robert2014} introduce ABC as a way to simulate the posterior distribution $\beta$. However, simulating a data set is unfortunately non-trivial for Markov random fields, as it usually requires a certain number of steps of an MCMC sampler.

It seems quite awkward to define the corresponding constraints that are needed to use an empirical likelihood procedure; therefore we will make the comparison of BC$_{bl}$ with an ABC approach.
We compare the performance of ABC and $\mbox{BC}_{bl}$ in a simulation dataset of size $25\times25$ where the true parameter $\beta$ is set as 0.5. The simulation is done using the Gibbs sampler, starting with a random configuration with each pixel being drawn independently from $\{0,1\}$, and then iterating for 200 Gibbs cycles. The sufficient statistic $S$ is
\begin{equation}
S(\bm{x})=\sum_{(i,j)\sim (i',j')}\mathbbm{l}_{\{x_{ij}=x_{i'j'}\}}. \label{eq.Sx}
\end{equation}

In order to preserve the spatial structure of data we consider blocks of pixels as bootstrap sampling units, and we apply moving block bootstrap (MBB) methods as suggested by \cite{Lahiri03}. A simulation study about optimum block dimensions can be found in \cite{Zhu04}. 

Then, in the simulated data, as the corresponding structure is a square grid of pixels, we use a square moving window of length side equal to 5, as it renders good performance to estimate the original parameters. Also, we use the MPLE technique for estimating the parameters in each iteration.
The numbers of bootstrap replicates for the 1st level and 2nd level bootstrap are 100 and 200, respectively. A $U(0,2)$ prior distribution is assumed for the parameter $\beta$. The choice of the interval $[0,2]$ is motivated by the critical value $\beta=1$ that represents the phase transition of the Ising Model.
Regarding the tuning parameters of the ABC approach, we use 20000 iterations and a tolerance level of 0.01. As summary statistics, we take the sufficient statistic $S$ which is defined in \eqref{eq.Sx}.

\begin{figure}[H] 
\centering
\includegraphics[width=10cm,height=8cm]{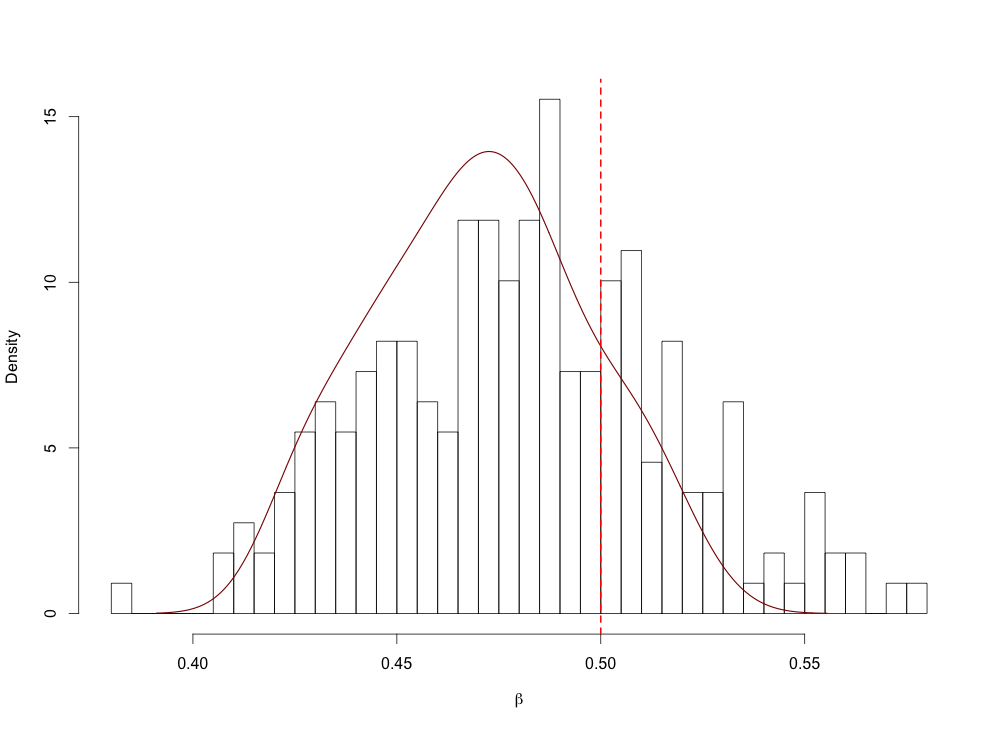}
\caption{\footnotesize{Comparison of the $\mbox{BC}_{bl}$ (curve) with the histogram of the simulations from ABC algorithm with $10^4$ iterations and a 1\% quantile on the difference between the sufficient statistics as its tolerance bound $\epsilon$, based on the uniform prior $U(0,2)$. \label{Ising1}}}
\end{figure}

Figure \ref{Ising1} shows that the estimation carried with $\mbox{BC}_{bl}$ and ABC algorithms provides similar results. It is worth to mention that the  $\mbox{BC}_{bl}$ has a  computational cost which is less than ABC since the Gibbs sampling for the Ising model has a cost which increases quadratically as the lattice grows (In our experiment, $BC_{bl}$ takes 3 hours and ABC takes 25 hours). The same problem arises with Potts model where more than two colors are considered.

We conclude this section with a real data example. We consider a set of soil phosphate measurements collected during the Laconia Archaeological Survey in Greece (year 1987).  A complete description of data can be found i.e. in \cite{Buck88}.
This dataset has been analysed by using different techniques, for instance \cite{Buck88} carried a Bayesian change-point analysis to describe the dataset. \cite{Besag1991} adopted a Bayesian image analysis approach. Recently,  \cite{McGrory2009} studied the dataset with variational Bayes methods.

\begin{figure}[H] 
\centering
\includegraphics[width=10cm,height=8cm]{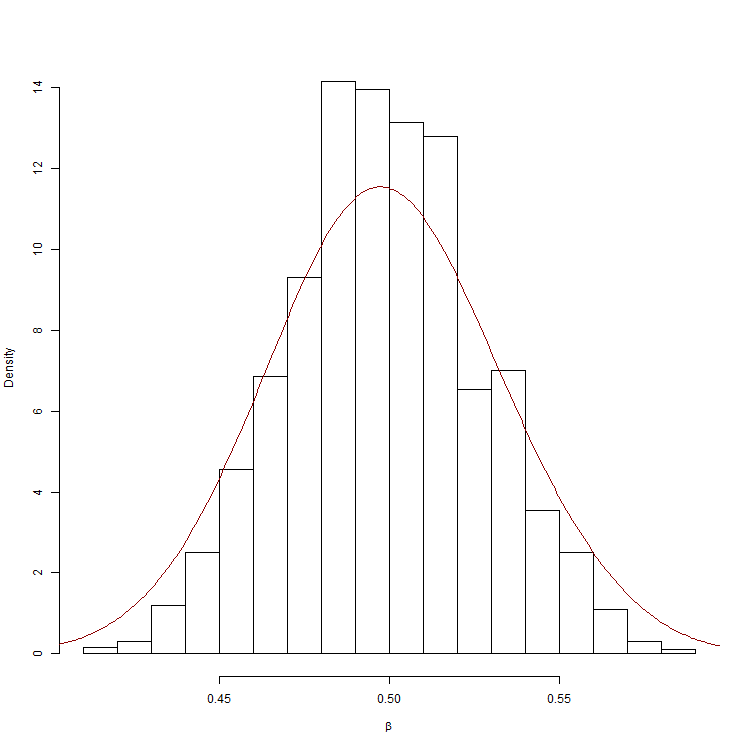}
\caption{\footnotesize{Histogram and density estimation of parameter $\beta$ after applying $\mbox{BC}_{bl}$ in the Laconia Archaeological data. \label{Isingreal}}}
\end{figure}

In this application, we use the moving block bootstrap and MPLE techniques in a similar way as in the simulation study. The window length of the moving block is set as 8. The numbers of bootstrap replicates for the 1st level and 2nd level bootstrap are 250 and 500, respectively. The distribution of values of $\beta$ is shown in Figure \ref{Isingreal}. 

The distribution of parameter $\beta$ is roughly located between 0.45 and 0.55; it may be noted that results are quite similar to those obtained in \cite{McGrory2009} who use a variational Bayes method. In their case, the estimation of parameter $\beta$, by using variational Bayes and MCMC methods, also renders similar estimates between 0.44 and 0.59 (see Table 3 of \cite{McGrory2009}).

\section{Discussion}


In this Section we discuss the performance of the algorithms and suggest possible improvements for future research. In particular, Table \ref{TabCTime} shows the computing times for the examples studied in Section \ref{Example}.

\begin{table}[h] \small
\centering
\begin{tabular}{|l|l|l|l|}
	\hline
	\textbf{Models}      & BC$_{bl}$ & BC$_{el}$ & ABC      \\
	\hline
	\emph{GARCH}         & 1.3 min.   & 2.5 min.   & 2.8 min.  \\
	\hline
	\emph{Pop. Genetics} & 10 hours  & 4.5 hours & N/A      \\
	\hline
	\emph{SDE}           & 7 min.     & N/A       & 6 min.    \\
	\hline
	\emph{Ising Potts}   & 3 hours   & N/A       & 25 hours \\
	\hline
\end{tabular}
\caption{\footnotesize{Resume of computing times for the corresponding models.}}\label{TabCTime}
\label{table:times}
\end{table}

\noindent As anticipated in the population genetics example (Section \ref{PopGen}), the computing time of $\mbox{BC}_{bl}$ is more than twice the computing time of $\mbox{BC}_{el}$. This is not surprising since a clear estimator is not available in this case. On the other hand, $\mbox{BC}_{bl}$ is faster in the GARCH and Random Field  examples where well-known estimators of the parameters are available. The SDE example is quite interesting since the  $\mbox{BC}_{bl}$ exhibits a similar computing time when compared with the ABC, but with better estimates of the parameters. A final remark is about the computing time of the population genetics example. Table S1 of \cite{Robert13} reports times in seconds whilst we report times in hours (see Table \ref{TabCTime}). The reason is due to the different codification method; in particular, we use R and not C++ as in \cite{Robert13}. This suggests that future research should be devoted to efficient implementation of $\mbox{BC}_{bl}$. In particular, we think that $\mbox{BC}_{bl}$ could be efficiently parallelized, increasing the application of the method to more complex models. 

Finally, we devote the last part of this section to an idea that seems promising for future research. One of the anonymous reviewers suggested to perform the $\mbox{BC}_{bl}$ with an ABC estimator. 

\begin{table}[h]
\centering
\small
\begin{tabular}{|c|c|c|c|}
\hline
True values & ABC & $BC_{bl}$ (qmle) & $BC_{bl}$ (ABC)\\ \hline
$\theta_{1}=0.2$ & 0.28644 (0.01300) & 0.20144 (0.00008) & 0.2354 (0.00697)\\ \hline
$\theta_{2}=0.3$ & 0.41261 (0.02420) & 0.34773 (0.02360) & 0.33523 (0.02132)\\ \hline
\end{tabular}
\caption{\footnotesize{The first two columns report the results of Table \ref{table:SDE}. The last column reports the results of the $\mbox{BC}_{bl}$ with the ABC estimator of the first column.}}\label{BCBL_ABC}
\end{table}

We focused on the SDE example and we run $\mbox{BC}_{bl}$ 10 times, with 50 replicates in the first level bootstrap and 200 replicates in the second level bootstrap. The last column of Table \ref{BCBL_ABC} displays the results of the new simulation. Compared with the results of Section \ref{SDESEC}, which for convenience are reported in the first two columns of Table \ref{BCBL_ABC}, it is interesting to note that $\mbox{BC}_{bl}$ improves the performance of the ABC estimation. Although from the computational point of view, the procedure is expensive, the inferential results justify investment in future research about the use of an ABC estimator in a Bootstrap likelihood setting.

\section*{Acknowledgements}
The authors acknowledge the helpful comments of two anonymous referees which led to a significant improvement of the paper. The authors are very grateful to Pierre Pudlo for all the suggestions about the population genetic example. J. Miguel Marin's research has been supported by grant MTM2010-17323. Fabrizio Leisen's research has been supported by the Royal Society Grant "Bayesian Computational methods with Bootstrap Techniques" and by the European Community's Seventh  Framework Programme[FP7/2007-2013] under grant agreement no:630677.  


\bibliographystyle{elsarticle-harv}

\begin{thebibliography}{00}




\bibitem[Beaumont, Zhang and Balding (2002)]{Beaumont02} Beaumont, M., Zhang, W. and Balding, D. (2002). Approximate Bayesian computation in population genetics. Genetics 162(4), 2025--2035.

\bibitem[Besag (1977)]{Besag1977}
Besag, J. (1977). Efficiency of pseudolikelihood estimation for simple Gaussian fields. Biometrika 64(3), 616--618.

\bibitem[Besag, York and Mollie\ (1991)]{Besag1991}
Besag, J., York, J. and Mollie, A. (1991). Bayesian image restoration, with two applications in spatial statistics (with discussion). Ann. Inst. Stat. Math. 43(1), 1--59.

\bibitem[Bollerslev (1986)]{Bollerslev86}
Bollerslev, T. (1986). Generalized Autorregressive Conditional Heteroskedasticity. J. Econometrics 31(3), 307--327.

\bibitem[Brouste et al.\ (2014)]{Brouste14}
Brouste, A., Fukasawa, M., Hino, H., Iacus, S.M., Kamatani, K., Koike, Y., Masuda, H., Nomura, R., Ogihara, T., Shimuzu, Y., Uchida, M. and Yoshida, N. (2014). The YUIMA Project: A Computational Framework for Simulation and   Inference of Stochastic Differential Equations. J. Stat. Software 57(4), 1--51.

\bibitem[Buck, Cavanagh and Litton (1988)]{Buck88}
Buck, C.E., Cavanagh, W.G. and Litton, C.D. (1988). The spatial analysis of site phosphate data. In: Rhatz, S.P.Q. (ed.) Computer Applications and Quantitive Methods in Archeology. British Archaeological Reports, International Series, vol. 446. BAR, Oxford.

\bibitem[Cornuet et al.\ (2014)]{Cornuet14}
Cornuet, J.M., Pudlo, P., Veyssier, J., Dehne-Garcia. A., Gautier, M., Leblois, R., Marin, J.M. and Estoup, A. (2014). DIYABC v2.0: a software to make Approximate Bayesian Computation inferences about population history using Single Nucleotide Polymorphism, DNA sequence and microsatellite data. Bioinformatics 30(8), 1187--1189. 



\bibitem[Davison, Hinkley and Worton\ (1992)]{Davison92}
Davison, A.C., Hinkley, D.V. and Worton, B.J. (1992). Bootstrap likelihoods. Biometrika 79(1), 113--130.

\bibitem[Davison and Hinkley (1997)]{Davison97} Davison, A.C. and Hinkley, D.V. (1997). Bootstrap methods and their application. Cambridge University Press.

\bibitem[Davidson and  MacKinnon (2006)]{Davison06}
Davidson, R. and  MacKinnon, J.G. (2006) Bootstrap methods in econometrics. Palgrave Handbooks of Econometrics, ed. by TC Mills, and KD Patterson, 812--838.

\bibitem[Dean et al.\ (2014)]{Dean14} Dean, T., Singh, S., Jasra, A. and Peters, G. (2014). Estimation of HMMs with Intractable Likelihoods. Scand. J. Statist. 41(4), 970--987.

\bibitem[Drovandi and Pettitt (2010)]{Drovandi10} Drovandi, C. and Pettitt, A. (2010). Estimation of parameters for macroparasite population evolution using approximate Bayesian computation. Biometrics 67(1), 225--233.

\bibitem[Fearnhead and Prangle (2012)]{Fearnhead12} Fearnhead, P. and Prangle, D. (2012). Constructing summary statistics for approximate Bayesian computation: semi-automatic approximate Bayesian computation. J. Royal Statistical Society (B) 74(3), 419--474.

\bibitem[Grimmett (2010)]{Grimmett10}
Grimmett, G. (2010). Probability on Graphs Random Processes on Graphs and Lattices. Cambridge University Press.

\bibitem[Lahiri (2003)]{Lahiri03}
Lahiri, S.N. (2003). Resampling Methods for Dependent data.  Springer--Verlag, New York.

\bibitem[MacKinnon (2006)]{Mackinnon06}
MacKinnon, J. G. (2006). Bootstrap methods in econometrics. Economic Record, 82(s1), S2--S18.

\bibitem[Marin and Robert (2014)]{Robert2014}
Marin, J.M. and Robert, C.P. (2014). Bayesian Essentials with R. Springer--Verlag, New York. 

\bibitem[Marin et al (2012)]{Robert2012}
Marin, J.M., Pudlo, P., Robert, C.P., Ryder R. J. (2012).
Statistics and Computing 22(6), 1167--1180.

\bibitem[McGrory et al.\ (2009)]{McGrory2009}
McGrory, C.A.,  Titterington, D.M., Reeves, R. and Pettitt, A.N. (2009). Variational Bayes for estimating the parameters of a hidden Potts model. Stat. Comput. 19(3), 329--340.

\bibitem[McKinley, Cook and Deardon (2009)]{McKinley09} McKinley, T., Cook, A. and  Deardon, R. (2009). Inference in epidemic models without likelihoods. Int. J. Biostat. 5(1), 1--40.

\bibitem[Mengersen, Pudlo and Robert (2013)]{Robert13}
Mengersen, K.L., Pudlo, P. and Robert, C.P. (2013). Bayesian computation via empirical likelihood. Proceedings of the National Academy of Sciences 110(4), 1321--1326.

\bibitem[Del Moral, Doucet and Jasra (2012)]{Moral12}
Del Moral, P., Doucet, A. and  Jasra, A. (2012). An adaptive sequential Monte Carlo method for approximate Bayesian computation. Stat. Comput. 22(5), 1009--1020.

\bibitem[Olkin and Liu (2003)]{Olkin2003}
Olkin, I. and Liu, R. (2003). A bivariate beta distribution. Stat. Prob. Letters 62(4), 407--412.

\bibitem[Owen (2001)]{Owen01}
Owen, A.B. (2001). Empirical likelihood. CRC press.

\bibitem[Picchini (2014)]{picchi14}
Picchini, U. (2014). Inference for SDE Models via Approximate Bayesian Computation. J. of Comp. and Graph. Stat. 23(4), 1080--1100.

\bibitem[Pritchard et al. (1999)]{Pritchard} Pritchard, J., Seielstad, M., Perez-Lezaun, A. and Feldman, M. (1999). Population growth of human Y chromosomes: a study of Y chromosome microsatellites. Molecular Biology and Evolution 16(12), 1791--1798

\bibitem[Sisson, Fan and Tanaka\ (2007)]{Sisson07} Sisson, S.A., Fan, Y. and Tanaka, M. (2007). Sequential Monte Carlo without likelihoods. Proc. Natl. Acad. Sci. 104(6), 1760--1765.

\bibitem[Wilson and Balding (1998)]{Wilson98}
Wilson, I.J. and Balding, D.J. (1998). Genealogical inference from microsatellite data. Genetics 150(1), 499--510.

\bibitem[Xu and Reid (2011)]{XR2011}
Xu, X. and Reid, N. (2011). On the robustness of maximum composite likelihood estimate. Journal of Statistical Planning and Inference 149(9), 3047--3054.

\bibitem[Lin and Wang (2010)]{Zhengyan10} Lin, Z. and Wang, H. (2010). Empirical likelihood inference for diffusion processes with jumps. Science China. 53( 7), 1805-–1816.

\bibitem[Zhu and Morgan (2004)]{Zhu04} Zhu, J. and Morgan, G.D. (2004). Comparison of Spatial Variables over Subregions Using a Block Bootstrap. J. of Agricultural, Biological, and Environmental Statistics. 9(1), 91--104.

\end{thebibliography}


\end{document}